\documentclass[
 reprint,
superscriptaddress,
showpacs,preprintnumbers,
 amsmath,amssymb,
 aps,
 prl,
]{revtex4-1}

\usepackage{graphicx}
\usepackage{dcolumn}
\usepackage{bm}
\usepackage{hyperref}

\begin{document}


\title{Temperature dependent infrared spectroscopy of the Rashba spin-splitting semiconductor BiTeI}

\author{C. Martin}
\affiliation{Department of Physics, University of Florida, Gainesville, Florida 32611, USA}
\author{K. H. Miller}
\affiliation{Department of Physics, University of Florida, Gainesville, Florida 32611, USA}
\author{S. Buvaev}
\affiliation{Department of Physics, University of Florida, Gainesville, Florida 32611, USA}
\author{H. Berger}
\affiliation{Institute of Physics of Complex Matter, Ecole Polytechnique Federal de Lausanne, CH-1015 Lausanne, Switzerland}
\author{X.~S.~Xu}
\affiliation{Materials Science and Technology Division, Oak Ridge National Laboratory, Oak Ridge, Tennessee 37831, USA}
\author{A. F. Hebard}
\affiliation{Department of Physics, University of Florida, Gainesville, Florida 32611, USA}
\author{D. B. Tanner}
\affiliation{Department of Physics, University of Florida, Gainesville, Florida 32611, USA}

\date{\today}

\begin{abstract}
We performed temperature dependent infrared spectroscopy measurements on BiTeI single crystals, which exhibit large Rashba spin-splitting. Similar to a previous optical study, we found electronic excitations in good agreement with spin-split electronic bands. In addition, we report a low energy intraband transition with an onset energy of about 40 meV and an unexpectedly large number of vibrational modes in the far-infrared spectral region. At least some of the modes have asymmetric Fano line-shape. These new observations cannot be explained considering only the bulk band structure or crystal symmetry of BiTeI, and we proposed that the optical response is also affected by the surface topology.
\end{abstract}
\pacs{74.25.Ha, 74.78.-w, 78.20.-e, 78.30.-j}
\maketitle
Exploring the spin component of the conduction carriers (spintronics) represents an active research field, with the goal of realizing faster, more complex semiconducting devices. One avenue is to combine a large spin orbit interaction (SOI) with lack of inversion symmetry, which will lift the spin degeneracy of the electronic bands. This could be realized either in bulk semiconductors that contain elements with large atomic number (i.e. large SOI) and that do not have a spatial inversion symmetry $I$~\cite{Rashba60}, or in 2D electron gases under an electric field applied perpendicular to the surface~\cite{Bychkov84}. The so called Rashba effect results in a splitting of the two spin branches of an electronic band in momentum space by a wavevector $k_{R}$, as illustrated in the inset of Fig.~\ref{Fig4}. The splitting energy is $2E_{R}=\alpha_{R}k_{R}$, where $\alpha_{R}$ is the Rashba parameter, a measure of the spin-splitting strength. 

For potential applications, it is desirable to obtain a large momentum splitting and a large Rashba parameter, i.e. splitting energy and highest values were initially reported at the asymmetric interfaces between Bi and noble metals~\cite{Koroteev04, Ast07}. However, recent work combining angle resolved photoemission (ARPES) and band structure calculations reported a large bulk Rashba splitting, with $\alpha\approx$~3.8 eV$\AA$, in the polar semiconductor BiTeI~\cite{Ishizaka11}. Subsequently, theoretical calculations and experimental measurements of optical conductivity confirmed that the electronic excitations in BiTeI are in good agreement with the bulk Rashba split electronic bands~\cite{Lee11}. Further ARPES studies suggested tridimensional character of the bulk bands~\cite{Sakano12, Landolt12}, but some also reported the existence of two-dimensional surface branches, possibly with even higher Rashba spin splitting~\cite{Landolt12,Crepaldi12}. Theoretical calculations on BiTeX (X=Cl, Br, I) also indicate the reconstruction of the band structure at the Te-terminated surface, resulting in the formation of a surface 2D electron system that separates from the bulk states~\cite{Eremeev12}. Furthermore, we reported very recently the observation of quantum oscillations from surface carriers in BiTeI~\cite{Martin12}.

Previous optical studies, with the lower frequency limit of about 10 meV (80 cm$^{-1}$), focused on the optical response beyond the energy range dominated by the free carriers, concerning mostly with the intra and inter-band transitions between the Rashba split branches~\cite{Lee11}. Here we report measurements of temperature dependent infrared spectroscopy, from lower energy ($\approx$~3 meV), on single crystals of BiTeI. In addition to previous findings, we report new features in the optical spectrum: an electronic excitation with a low energy onset of about 40 meV and a complex vibrational spectrum, that cannot be explained from the symmetry of the bulk crystal structure.

Single crystals of BiTeI were grown by chemical vapor transport and Bridgman method. Samples with shiny, mirror-like surface and diameter 3-5 mm were selected from the same batch. Measurements of resistance and Hall coefficient were performed on two of the samples using a commercial PPMS with base temperature of 2K and magnetic field up to 6 Tesla. Infrared optical reflectance at frequencies from 25 cm$^{-1}$ to 6000 cm$^{-1}$ and at several temperatures between room temperature and 10 K, was measured using a helium flow cryostat mounted on a Bruker-113v FTIR spectrometer. The high frequency reflectance, up to 25 000 cm$^{-1}$ (3 eV) was measured at room temperature using a Carl Zeiss microscope photometer and all other temperatures were merged at high frequency with the room temperature data. Optical conductivity was obtained from Kramers-Kronig transformation of the reflectance.
\begin{figure}
\includegraphics[width=0.45\textwidth]{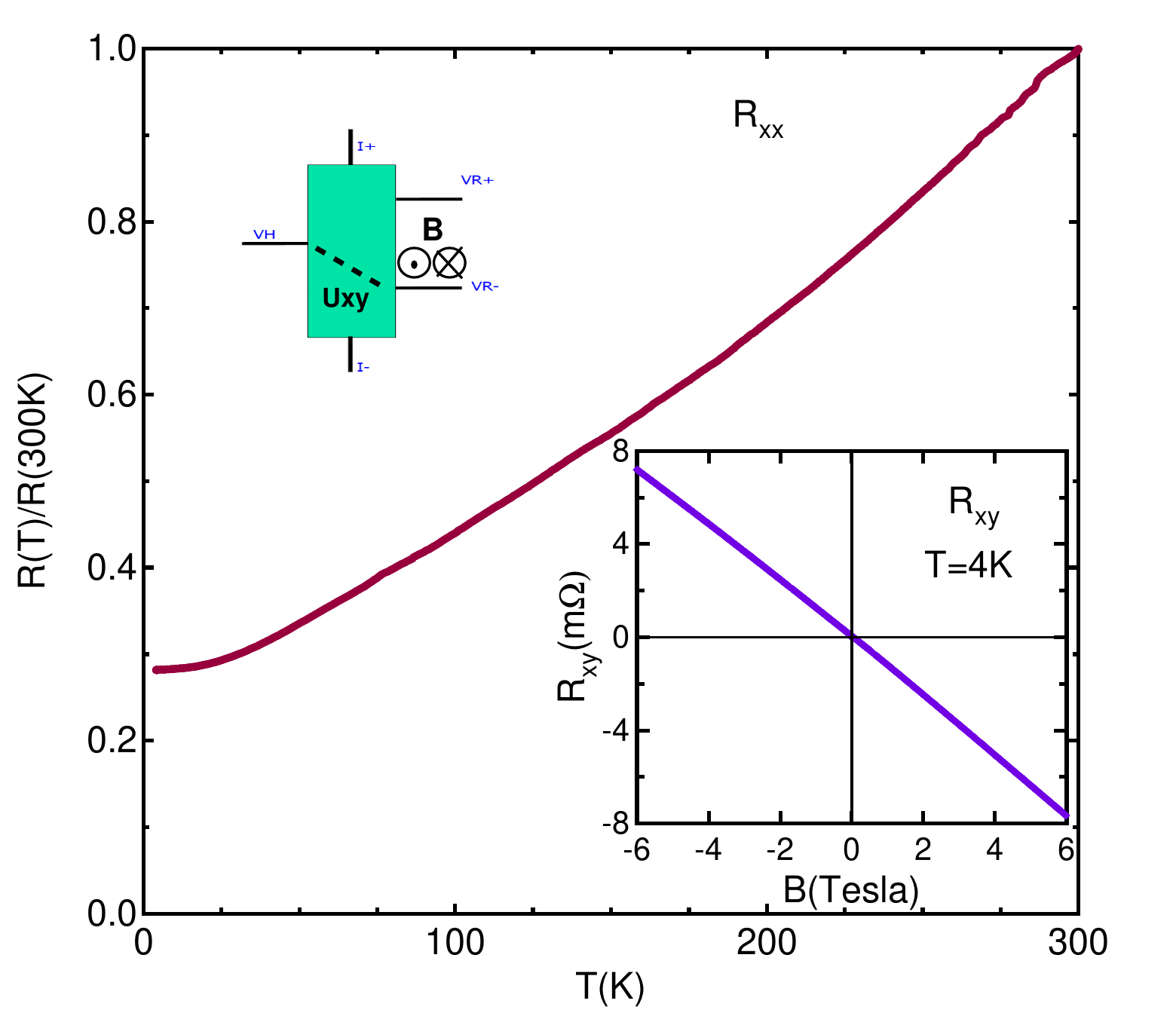}                
\caption{(Color online) Temperature dependence of BiTeI resistance normalized to room temperature value. Upper left inset: schematic diagram of the contacts configurations. Lower right inset: Hall resistance versus magnetic field at T = 4 K.} 
\label{Fig1}
\end{figure}

From the main panel of Fig.~\ref{Fig1} can be seen that the temperature dependence of resistance shows metallic behavior, decreasing by more than three times with cooling from room temperature to T= 2 K. Although BiTeI is a bulk semiconductor with  E$_{g}\approx$~0.4 eV,  resistance data suggests that the Fermi energy is situated within the conduction band, most likely because of slight non-stoichiometry of this compound~\cite{Tomokiyo77}. The Hall resistance shown in the lower inset of Fig.~\ref{Fig1} indicates that the free carriers are electrons and we estimated an electron concentration $n_{e}\approx$ 5$\times$10$^{19}$ cm$^{-3}$. Measurements on two separate samples yielded very similar results, suggesting consistency in the quality of the samples from the batch.

\begin{figure}
\includegraphics[width=0.45\textwidth]{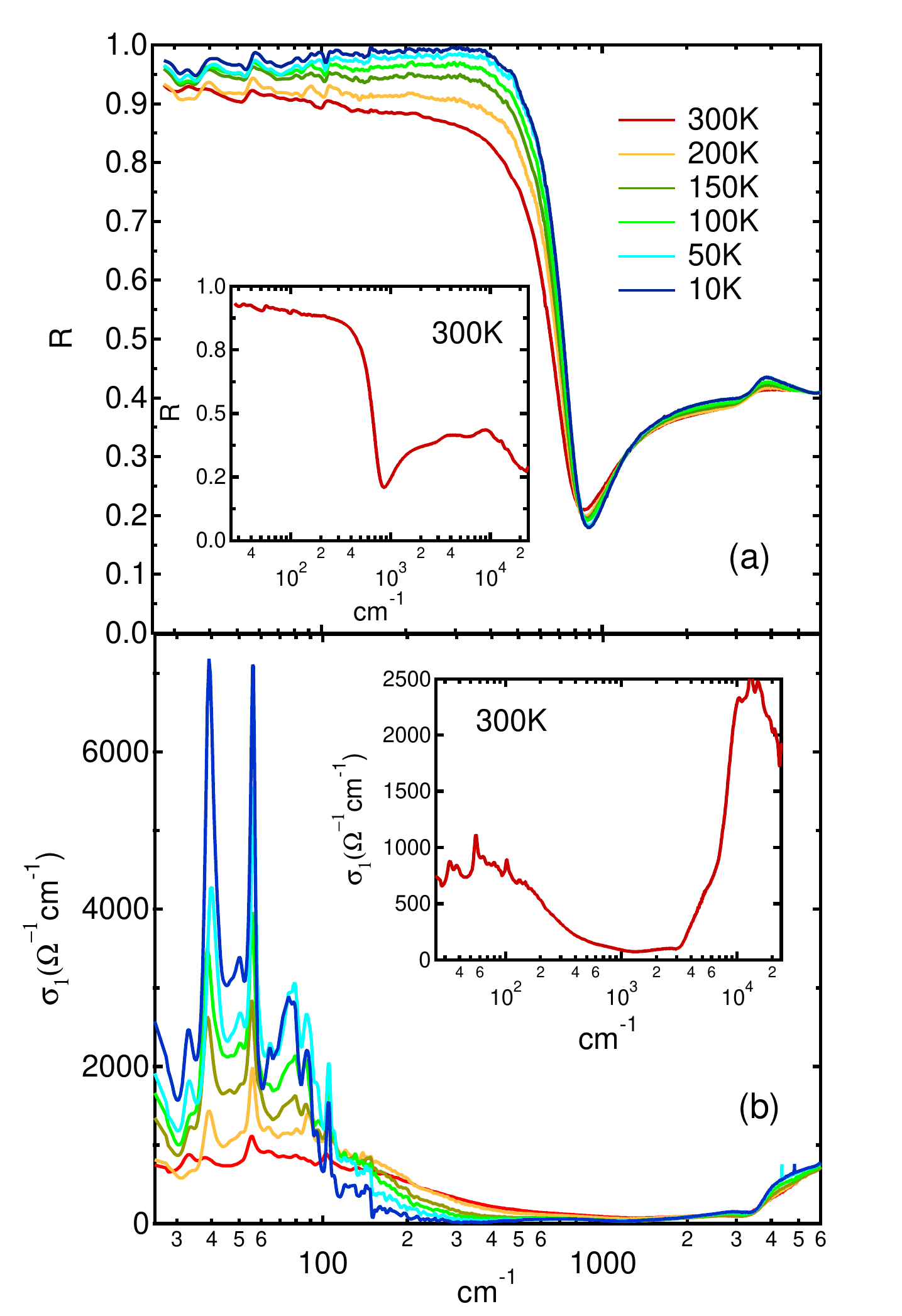}                
\caption{(Color online) Temperature dependent infrared reflectance R($\omega$) (a) and real part of optical conductivity $\sigma_{1}(\omega)$ (b). Insets shows room temperature R($\omega$) (a) and $\sigma_{1}(\omega)$ (b) up to 25 000 cm$^{-1}$.} 
\label{Fig2}
\end{figure}

Figure~\ref{Fig2}(a) shows the infrared reflectance, $R(\omega)$, for different temperatures between 10 K and 300K.
At low frequency, reflectance is higher than 90$\%$ and increases with cooling, consistent with the metallic character of the resistivity from Fig~\ref{Fig1}. Several sharp features can be observed below 200 cm$^{-1}$ and they will be discussed later. The reflectance minimum at 860 cm$^{-1}$ ($\approx$0.1 eV) sharpens and shifts slightly to higher frequency (by about 5 meV) at low temperature. Real part of optical conductivity, $\sigma_{1}(\omega)$, obtained from Kramers-Kronig transformation is shown in Fig.~\ref{Fig2}(b). The low frequency spectrum consists of a peak centered at zero frequency (Drude peak), due to the free carriers and multiple phonon modes associated with lattice vibrations. The temperature dependence of the zero frequency extrapolation $\sigma_{DC}$(T) is consistent with resistivity data, increasing at 10 K by more than three times from its room temperature value. Beyond the width of the Drude contribution, $\sigma_{1}(\omega)$ nearly vanishes up to 3100 cm$^{-1}$ (0.38 eV), where a clear edge marking the onset of interband transitions can be observed, particularly in the inset of  Fig.~\ref{Fig2}(b). The value of the edge is in excellent agreement with that of the bulk semiconducting gap determined previously from ARPES~\cite{Ishizaka11} and optical measurements~\cite{Lee11}. 

\begin{figure}
\includegraphics[width=0.45\textwidth]{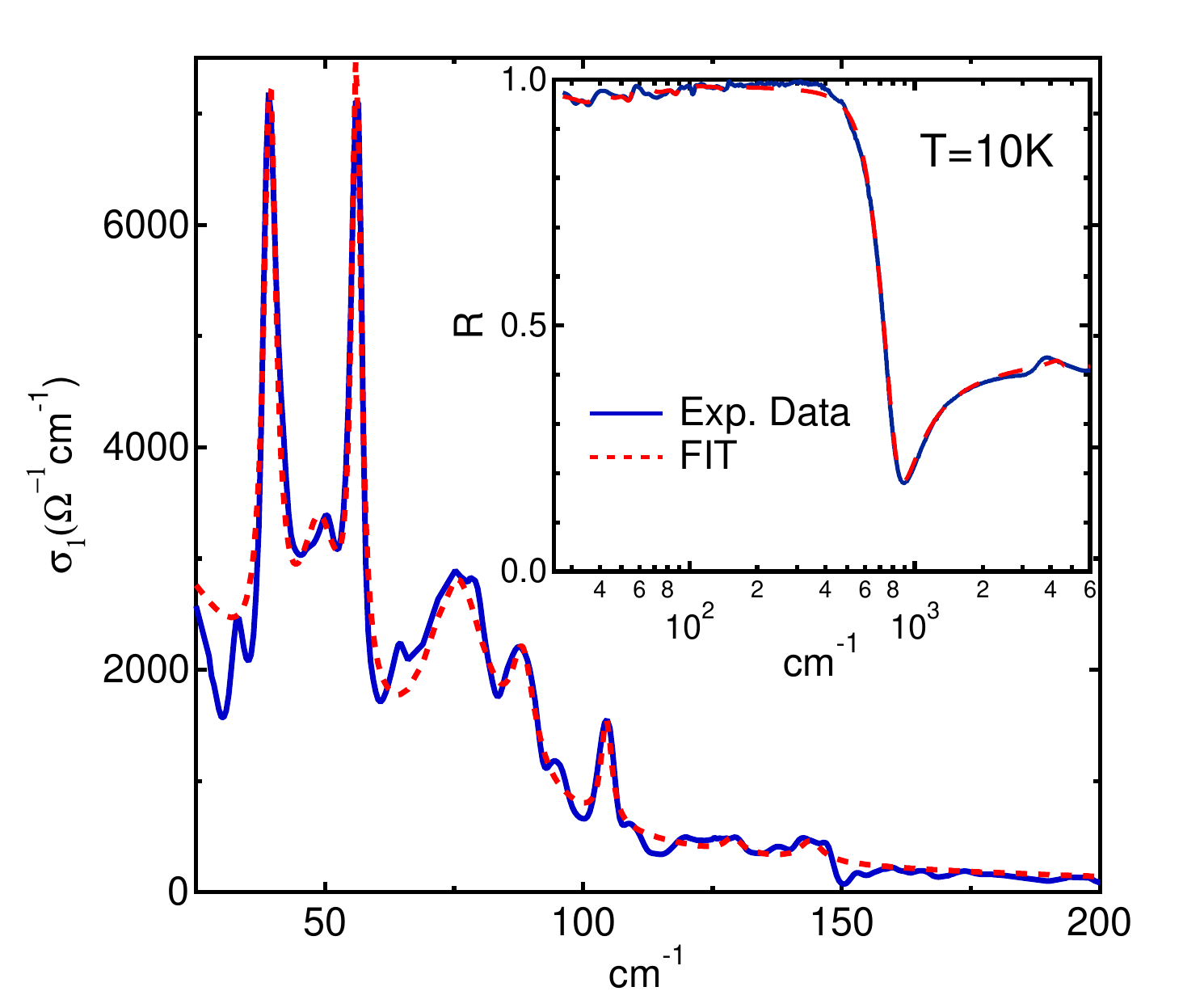}                
\caption{(Color online) Lorentz-Drude (LD) fit of the far-infrared region of optical conductivity at T=10 K. Inset shows the fit of R($\omega$) over the full spectral range.} 
\label{Fig3}
\end{figure}

\begin{figure}
\includegraphics[width=0.45\textwidth]{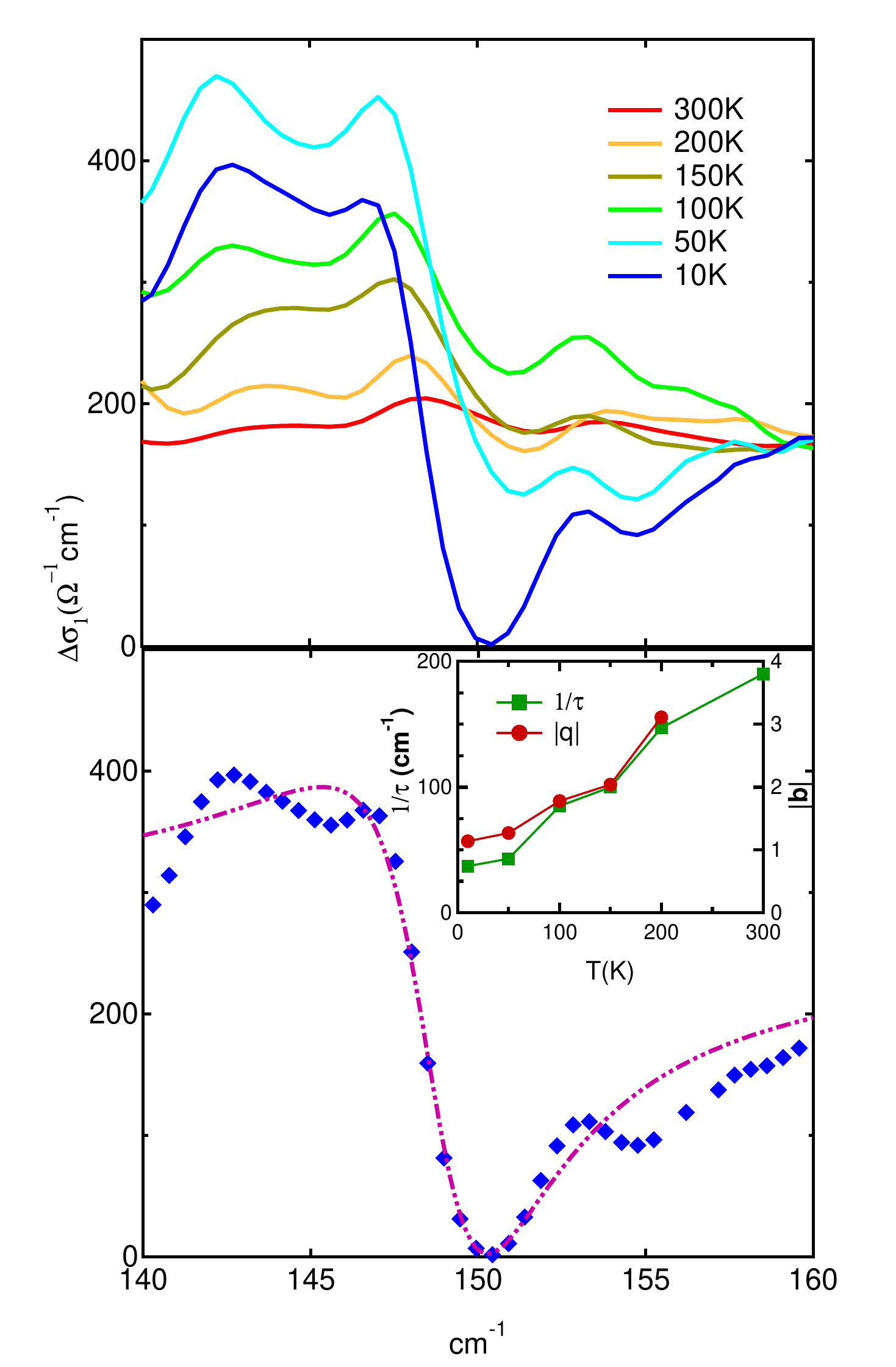}                
\caption{(Color online) (a) $\Delta\sigma_{1}$ around the phonon mode at 150 cm$^{-1}$ for different temperatures, after subtracting the Drude background. Traces were shifted vertically for clarity. (b) $\Delta\sigma_{1}$ at T= 10K (symbols) and fit to Eq.~\ref{Eq1} (dashed line). Inset displays on the same graph the temperature dependence of the asymmetry parameter $q$, obtained from the fit to Eq.~\ref{Eq1} of $\Delta\sigma_{1}$, and of the Drude scattering rate 1/$\tau_{D}$, from the LD fit of $\sigma_{1}$ and R.}   
\label{Fig4}
\end{figure}

We first discuss in more detail the low frequency behavior of optical conductivity. As can be seen from Fig.~\ref{Fig2}(b), there is large contribution to $\sigma_{1}(\omega)$ from the phonon modes, which overlap with the free carrier (Drude) response. For better accuracy in determining the Drude contribution (zero frequency peak), we fit both optical conductivity and reflectance, for each temperature, using a Lorentz-Drude (LD) model. Main panel of Fig.~\ref{Fig3} shows the low frequency region of the fit for $\sigma_{1}(\omega)$ at 10 K and the inset shows the result of the fit for the entire measured range of the reflectance, using the same fitting parameters as for conductivity. We found that the same value for the Drude plasma frequency $\omega^{D}_{p}$=2900 cm$^{-1}$ may be used for all temperatures and the resulting scattering rate of the conduction electrons decreases from $1/\tau^{D}\approx$180 cm$^{-1}$ at T= 300 K to about 40 cm$^{-1}$ at T=10 K, as shown in the inset of Fig.~\ref{Fig4}. Based on the carrier concentration from the Hall resistance data and on the value of  $\omega^{D}_{p}=\sqrt{4\pi~n_{e}e^{2}/m^{*}}$ from the Lorentz-Drude fits, we estimate an in-plane effective mass $m^{*}\approx$ 0.5$m_{e}$ in our samples. It is worth mentioning that in our recent study, we concluded that the optical response in BiTeI is dominated by the bulk carries, therefore the parameters obtained above correspond to the bulk~\cite{Martin12}.
The parameters for the vibrational modes obtained from the fit in Fig.~\ref{Fig3} are summarized in Table~\ref{Phonons}. We identified unambiguously seven modes and we confirmed them by measuring two other samples. Besides, we performed measurements on both as-grown and fresh-cleaved surfaces and found the same results, therefore it is unlikely that they originate from surface contamination. Polarization dependent reflectance, with {\bf E} rotating every 30$^{\circ}$ up to 120$^{\circ}$, was also measured, albeit only at room temperature, and it was found that the modes are isotropic. 
\begin{table}[ht]
\caption{In plane vibrational spectrum of BiTeI at T=10K} 
\centering 
\begin{tabular}{c c c c} 
\hline\hline 
Index & Plasma Freq. & Center Freq. & Width\\ [0.65ex] 
 & $\omega_{P}$ (cm$^{-1}$) & $\omega_{0}$ (cm$^{-1}$) & $\Gamma$ (cm$^{-1}$) \\
\hline 
1 & 975 & 39 & 3 \\ 
2 & 865 & 56 & 2.3 \\
3 & 1367 & 75.4 & 14 \\
4 & 596 & 88.3 & 5.4 \\
5 & 388.7 & 104.5 & 2.5 \\
6& 180 & 129 & 4\\
7 & 268 & 143 & 5.3 \\ [1ex] 
\hline 
\end{tabular}
\label{Phonons} 
\end{table}
We make two important observations about the phonon spectrum. First, their number is much larger than that expected from the crystal structure of BiTeI and from our configuration with the electric field along the $ab$-plane. BiTeI belongs to the $P3m1$ space group, without an inversions $I$-symmetry and with a $C_{3v}$ symmetry along the c-axis~\cite{Tomokiyo77, Shevelkov95}. From symmetry analysis, there should be a total of four vibrational modes, all IR-active. Two of them are a symmetric stretch and a symmetric deformation, respectively, both along the $c$-axis. The other two are a doubly-degenerate asymmetric stretch and a doubly-degenerate asymmetric deformation, respectively, both in the $ab$-plane. Therefore, only two modes are expected for $E\| ab$ and we cannot elucidate at this time the origin of the additional ones. However, their large number suggests the existence of a lower symmetry structure in BiTeI.

Our second observation is that, at least some of the vibrations have an asymmetric line shape. In fact, this asymmetry can be directly observed from the fit in Fig.~\ref{Fig3}, which reveals that the Lorentz-Drude model does not capture correctly the line-shape for some of the modes. While it may appear arguable for those up to 100 cm$^{-1}$, due to the close proximity of their center frequencies, it is obvious that the more isolated phonon mode at 143 cm$^{-1}$ is strongly asymmetric. We look more closely at this mode in Fig.~\ref{Fig4}(a), where we plot the optical conductivity $\Delta\sigma_{1}$, after subtracting the Drude background at each temperature. It appears more clear now that the mode strengthens, but also becomes more asymmetric at low temperature. We propose that the line shape is better described within the Fano physics~\cite{Fano61}, where the discrete vibrational energy of the mode is affected by a continuum electronic excitation. The contribution of the Fano resonance to optical conductivity $\Delta\sigma_{1}$ is given by~\cite{Fano61}:
\begin{equation}
\Delta\sigma_{1}(\omega)=\frac{\omega^{2}_{p,F}}{4\pi\Gamma}\frac{q^{2}+2qz-1}{q^{2}(1+z^{2})},
\label{Eq1}
\end{equation}
where $\omega_{p,F}$ is the plasma frequency of the Fano oscillation, $\Gamma$ is the line-width,$z=2(\omega-\omega_{0})/\Gamma$, $\omega_{0}$ being the center frequency, and $q$ is a parameter that defines the asymmetry of the resonance. The parameter $q$ is positive when the phonon interacts with a lower energy excitation and is negative when the excitation has higher energy. The resonance is more asymmetric for small absolute values of $q$ and recovers the symmetric Lorentz shape in the limit $|q|\rightarrow \infty$. Figure~\ref{Fig4}(b) shows the result of the fit to Eq.~\ref{Eq1} for $\Delta\sigma_{1}(\omega)$ at T= 10 K. We obtained a value of $q=-0.65\pm 0.06$, which indicates a strongly asymmetric line shape. The negative value suggest interaction of the mode with a higher energy continuum. 
\begin{figure}
\includegraphics[width=0.45\textwidth]{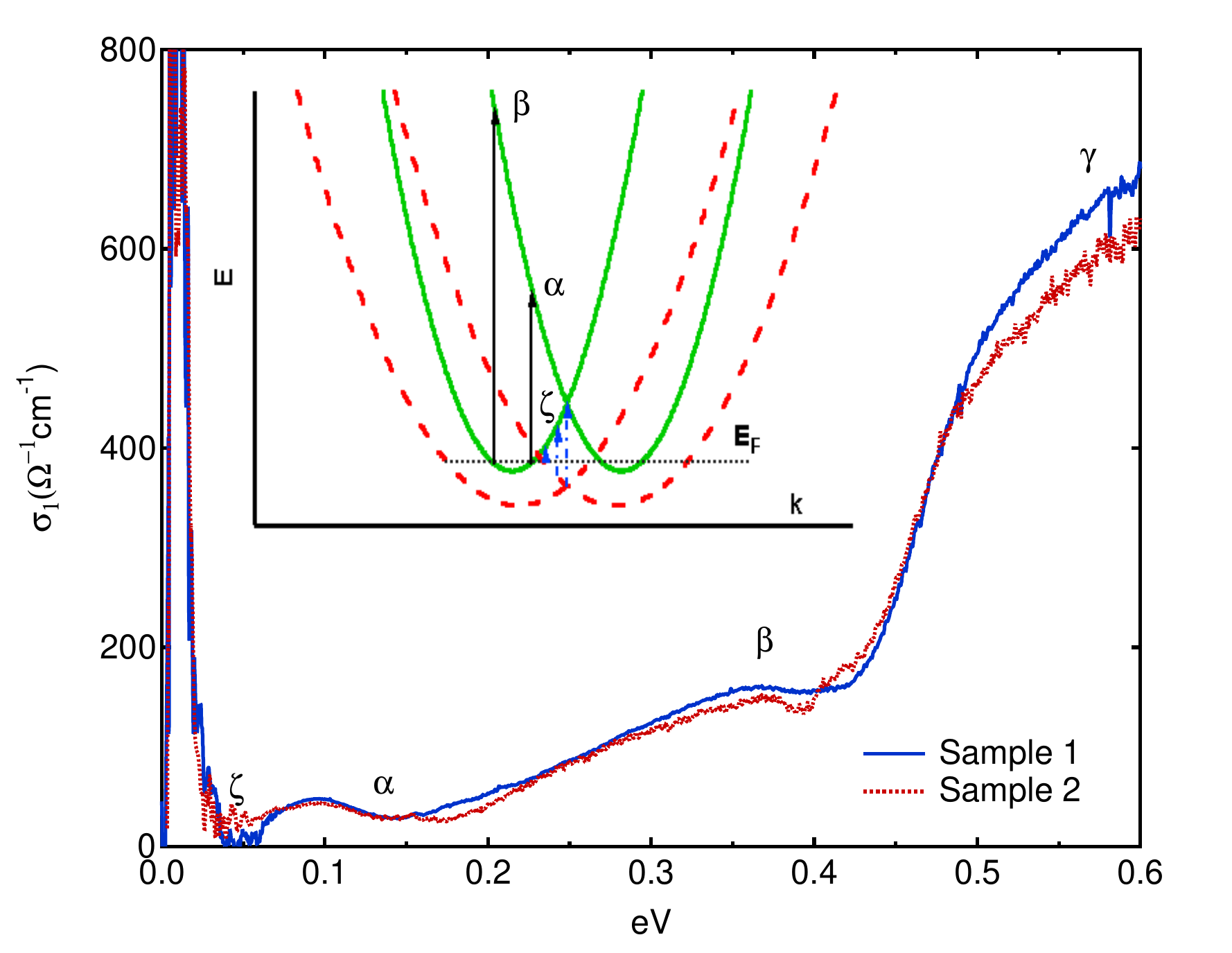}                
\caption{(Color online) Optical conductivity for two different samples at T=10 K, after subtracting the Drude background. Inset represents a sketch of conduction band in BiTeI, including both bulk (continuous lines) and possible surface branches (dashed lines).} 
\label{Fig5}
\end{figure}

Not having established the origin of the vibration, it is difficult to understand the mechanism for its Fano line-shape. We notice however, from the inset of Fig.~\ref{Fig4}(b), that there is a remarkable correlation between the temperature dependence of the asymmetry parameter $|q|$(T) and that of the Drude scattering rate $1/\tau^{D}$(T). This suggest the possibility of a coupling between the phonon and an electronic transition involving free carriers. The so called charged-phonon was proposed by Rice et al.~\cite{Rice77} to explain the line-shape asymmetry observed in organic conductors~\cite{Tanner77} and it was later on used for the analysis of the vibrational modes in other systems, like doped C$_{60}$~\cite{Rice92}, bilayer graphene~\cite{Kuzmenko09} and Bi-based topological insulators~\cite{LaForge10, Pietro12}.

An excitation of the conduction electrons that give rise to an electronic polarization of the space may couple to a vibrational mode, resulting in a spectral weight transfer from this transition to the phonon. The coupling constant, denoted $g$ in Ref.~\cite{Rice77}, will increase at low temperature as the free carriers scattering rate is reduced. Noticing that the contribution to the dielectric constant, and hence spectral weight transfer is proportional to $g^{2}$, it is therefore expected that the phonon mode will be stronger affected, therefore more asymmetric at low temperature, like we found in Fig.~\ref{Fig4}(b). We suggest that the electronic excitation involved in BiTeI is an intraband transition between bulk and surface conduction branches, as it will be discussed below.

We turn now to the behavior of optical conductivity at higher energy. Previous optical study has shown that the Rashba spin splitting of the electronic bands in BiTeI give rise to a more complex electronic excitation spectrum. Following the approach from Ref.~\cite{Lee11}, in Fig.~\ref{Fig5} we plot $\Delta\sigma_{1}(\omega)$ at T=10 K for two different samples, obtained after subtracting the Drude contribution. There is still finite optical conductivity at low frequency in our data, due to the phonon modes (not discussed in the previous study). Nevertheless, we identified all the intraband and interband transitions reported in Ref.~\cite{Lee11} and labeled $\alpha$, $\beta$ and $\gamma$. The higher energy interband transition $\delta$ from Ref.~\cite{Lee11} appears clearly in our room temperature data (see inset of Fig.~\ref{Fig2}(b)). Additionally, we observe from Fig.~\ref{Fig5} that there is another electronic excitation, centered at 0.1 eV with the onset $\zeta$ as low as 0.04 eV. We propose that this transition reveals an important detail in the band structure of BiTeI. Given its low energy scale, the onset $\zeta$ must represent an intraband transition. It was shown theoretically that the Rashba splitting of the bulk conduction band in BiTeI allows only the $\alpha$ and $\beta$ excitations, regardless of the position of the Fermi energy with respect to the band crossing point~\cite{Lee11}. Therefore, the transition $\zeta$ involves the existence of additional electronic branches. In light of the recent experimental data~\cite{Landolt12, Crepaldi12, Martin12} and $ab-initio$ calculations~\cite{Eremeev12}, that suggests the presence of surface states, we assign the additional band to a transition between the surface and the bulk states, like sketched in the inset of Fig.~\ref{Fig5}. 

In conclusion, we performed an infrared optical study on single crystals of BiTeI. We found a vibrational spectrum more complex than predicted by the crystal structure. Also, the phonon modes have strongly asymmetric line-shape, possibly due to their coupling to electronic excitations between bulk and surface states of the free carriers. This is supported by our observation of an additional intraband transition to those previously reported, which we assign to excitations between bulk and surface electronic branches.   

This work was supported by the U.S. Department of Energy through contract No.~DE-FG02-02ER45984 at the University of Florida.


\end{document}